\documentclass[a4paper]{jpconf}
\usepackage{graphicx}

\begin{document}
\newcommand{\hetre}{$^3$He } 
\newcommand{\cm}{\,{\rm cm}} 
\newcommand{\GeV}{\,{\rm GeV}} 
\newcommand{\MeV}{\,{\rm MeV}} 
\newcommand{\nanos}{\,{\rm ns}} 
\newcommand{\micros}{\,\mu{\rm s}} 
\newcommand{\millis}{\,{\rm ms}} 
\newcommand{\MSperSec}{\,{\rm MS}/{\rm s}} 

\title{NEUCAL, an innovative neutron detector for e/h discrimination: testbeam results}

\author{G~Sguazzoni$^{1}$, O~Adriani$^{2,1}$, L~Bonechi$^{2,1}$, M~Bongi$^{1}$, S~Bottai$^{1}$,
M~Calamai$^{3,1}$, G~Castellini$^{4}$, R~D'Alessandro$^{2,1}$, M~Grandi$^{1}$,
P~Papini$^{1}$, S~Ricciarini$^{1}$, P~Sona$^{2,1}$ and G~Sorichetti$^{2}$}

\address{$^1$ Istituto Nazionale di Fisica Nucleare, Sezione di Firenze, 50019, Sesto Fiorentino (FI), IT}
\address{$^2$ Dipartimento di Fisica, Universit\`a degli Studi di Firenze, 50019, Sesto Fiorentino (FI), IT}
\address{$^3$ Dipartimento di Fisica, Universit\`a degli Studi di Siena, 53100, Siena, IT}
\address{$^4$ IFAC-CNR Firenze, 50019, Sesto Fiorentino (FI), IT}

\ead{giacomo.sguazzoni@fi.infn.it}

\begin{abstract}
An excellent hadron to electron discrimination is a crucial aspect of calorimeter-based experiments in astroparticle physics.
Standard discrimination techniques require full shower development and fine granularity but in space detectors severe limitations exist due to constraints on dimensions, weight and power consumption.
A possible approach is to exploit the different neutron yield of electromagnetic and hadronic showers. NEUCAL is a light and compact innovative neutron detector, to be used as an auxiliary complement of electromagnetic calorimeters.
This new approach to neutron counting relies
on scintillation detectors which are sensitive to the moderation phase of the neutron component.
The NEUCAL prototype has been placed after a conventional calorimeter
and tested with high energy beams of pions and positrons.
The comparison of experimental data with a detailed Geant4 simulation
and the encouraging results obtained are presented.
\end{abstract}

\section{Introduction}

Space experiments designed to study cosmic ray fluxes at very high
energies must have, among other features, an excellent hadron to
electron discrimination. Electromagnetic calorimeters have a very good
intrinsic hadron rejection power based on the shower topology (depth
and lateral shape). Unfortunately, to be fully exploited, this
property requires calorimeters with a high granularity readout and
sufficient thickness to ensure
the full containment of the shower. 
These features cannot be fully implemented on satellites and airborne
experiments in general due to weight constraints and power consumption
limitations.
The hadron to electron discrimination capabilities are consequently reduced.

A workaround solution was attempted
in the successful PAMELA experiment~\cite{PAMELA}. A neutron detector, consisting of a moderator and a set of \hetre
counters, was placed downstream of the
calorimeter. The aim was to exploit the neutron component of the
hadronic showers, much larger than in electromagnetic showers. This
neutron counter, despite its very low efficiency, has been
successfully used by the PAMELA collaboration for systematic effects
evaluation and for measuring the calorimeter efficiency.

The NEUCAL project aims to further pursue and expand the use of
neutron detectors coupled to calorimeters by introducing the new
technique of {\em active moderation}~\cite{elba}, that consists in
detecting the signal of neutrons while their energy degrades within
hydrogen-rich scintillators. The most promising time interval for the
detection of the neutron moderation signal is between $\sim10\nanos$ to $\sim100\nanos$ after the shower core, when the
peak of the neutron flux arrives.
In the hadronic showers neutrons within this time window are produced 
by nuclear excitation processes and have energies of the order of $1\MeV$.

\section{NEUCAL prototype}
\label{sec:proto}

\begin{figure}[t]
\hspace{0.01\textwidth}%
\includegraphics[width=0.55\textwidth]{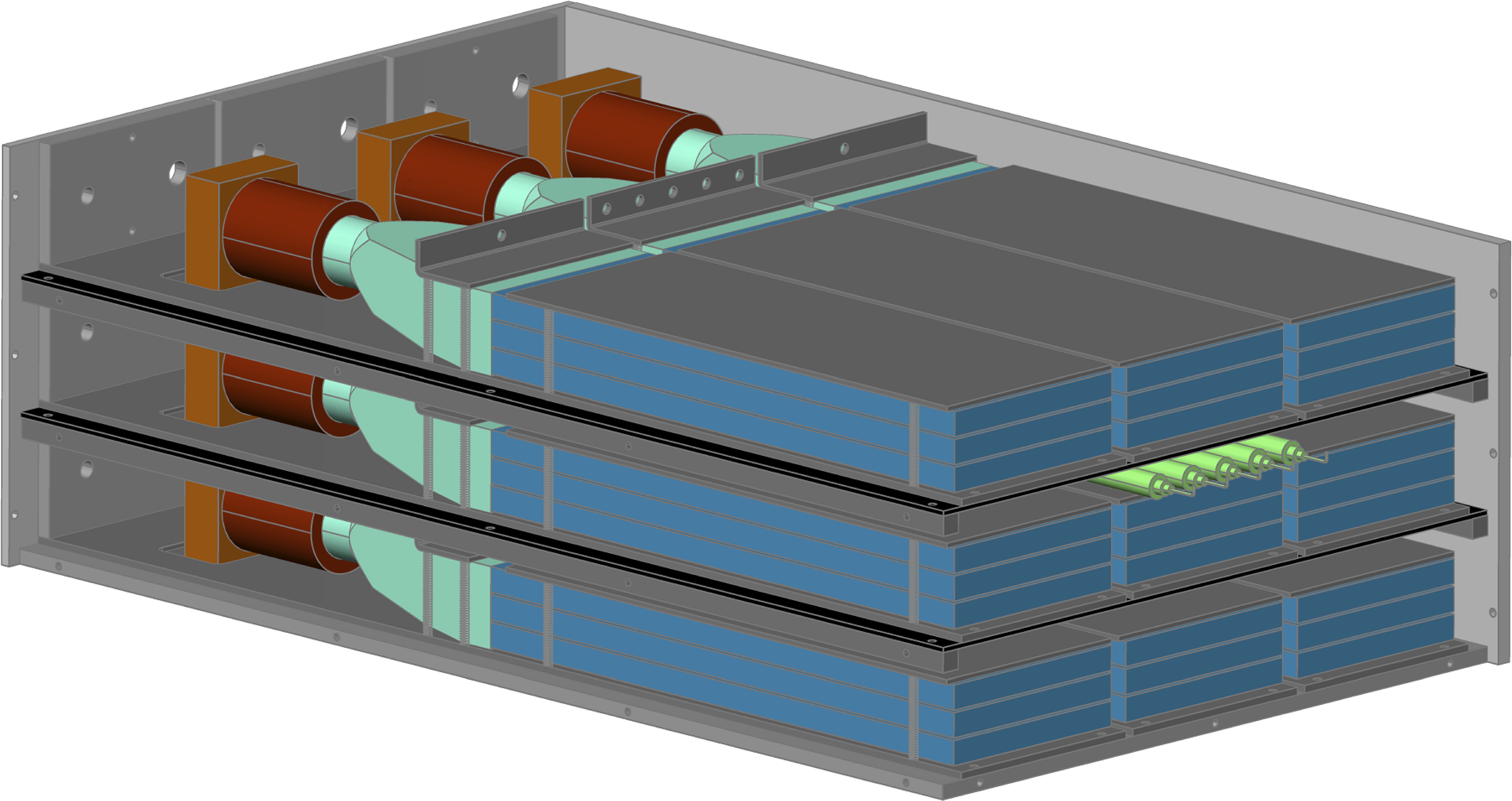}
\hspace{0.02\textwidth}%
\begin{minipage}[b]{0.41\textwidth}\caption{\label{fig:NeucalCAD}CAD open view of the NEUCAL detector prototype. The scintillators are in blue, and the five \hetre tubes (green) are placed on top of the central module.}
\end{minipage}
\end{figure}

The NEUCAL prototype~\cite{como, vienna}, sketched in Fig.~\ref{fig:NeucalCAD}, consists of a three by three matrix of identical modules placed in a light-tight aluminum box with three shelves. Each module is made up of three slabs ($25\cm\times8\cm\times1\cm$) of a fast polyvinyl-toluene scintillator (EJ-230 by Eljen Technology), coupled through a common Plexiglas light guide to a fast fine-mesh photomultiplier (R5946 by Hamamatsu Photonics). For the sake of flexibility, only optical grease was used in the optical couplings. Five \hetre tubes (12NH25/1 by Canberra) are also placed on top of the central module.
The scintillators and the tubes were read out by fast digitizers capable
of recording up to $10\millis$ of data samples. Two similar boards
were used: one CAEN V1731 with $500\MSperSec$ capability but only 8
bit range over an input dynamics of $1\,{\rm Vpp}$; one CAEN V1720 with a more limited $250\MSperSec$
capability but 12 bit range over an input dynamics of $2\,{\rm
  Vpp}$. Each board has eight input channels. For a better timing
precision one channel of each board was used to sample the trigger signal. Seven of the nine scintillator signals were sent to the $500\MSperSec$ board, while the two remaining scintillator signals and the five \hetre tubes signals were sent to the $250\MSperSec$ board.  The readout was performed with a VME system.

\section{Testbeam setup and simulation}

\begin{figure}[t]
\begin{minipage}[b]{0.49\textwidth}
\includegraphics[width=\textwidth]{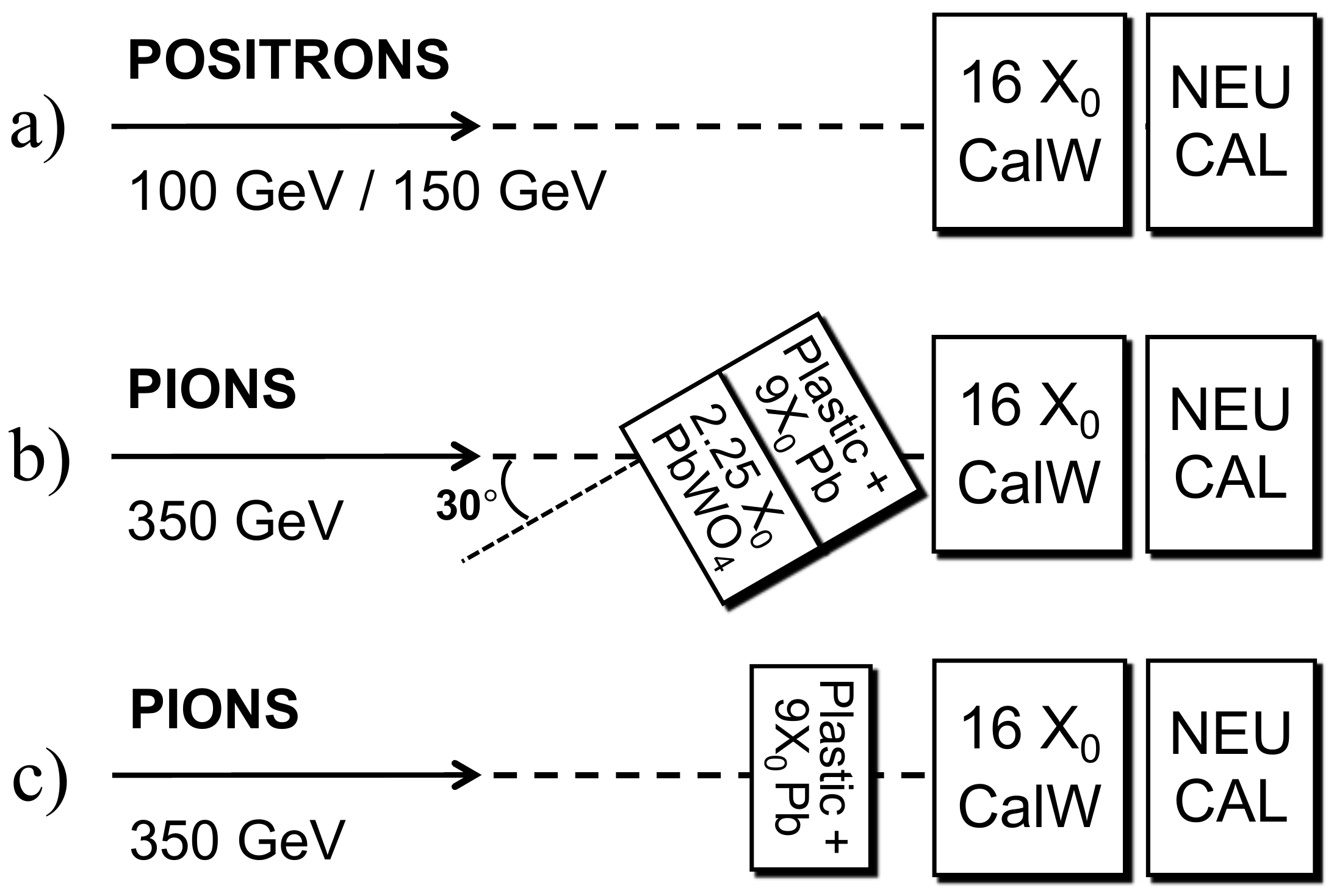}
\caption{\label{fig:TBConf}Sketch of the three testbeam configurations used for the present analyses of the NEUCAL data.}
\end{minipage}
\hspace{0.01\textwidth}%
\begin{minipage}[b]{0.49\textwidth}
\includegraphics[width=0.95\textwidth]{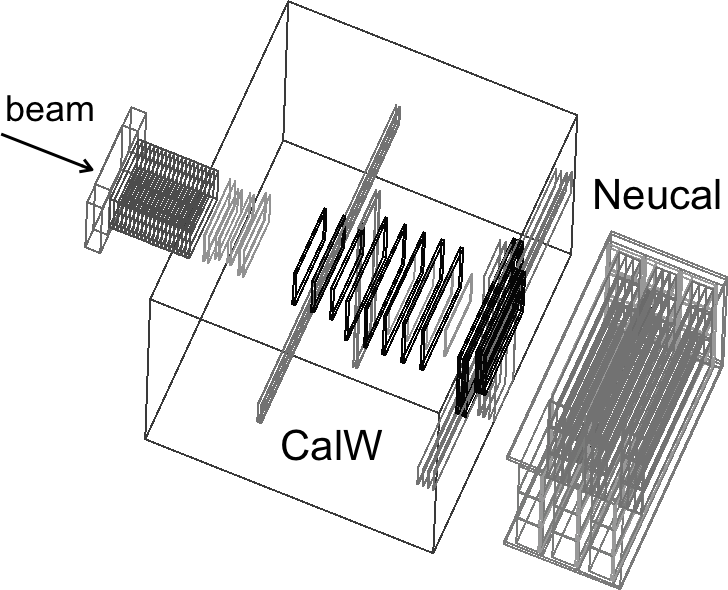}%
\caption{\label{fig:TBGeant}Testbeam geometry corresponding to configuration c) as implemented in Geant4.}
\end{minipage}
\end{figure}

In summer 2009 the NEUCAL prototype performances have been measured at CERN on the SPS line H4 during a test of the prototype of the the
Cream-2 tungsten-scintillator calorimeter ({\em CalW}\/)~\cite{cream2}. Data were collected for negative pions ($350\GeV$), positrons ($100\GeV$ and $150\GeV$), and muons ($150\GeV$)
which are not used for the present analysis.
In all testbeam configurations, sketched in Fig.~\ref{fig:TBConf}, NEUCAL was placed downstream CalW and all other possible devices and/or absorbers. In configuration a), used with positron beams, the upstream material corresponds to $16\,{\rm X_0}$ and $0.6\,\lambda_{\rm I}$; in configurations b) and c), used with pion beams, the upstream material corresponds to $29\,{\rm X_0}$ and $1\,\lambda_{\rm I}$, and $25\,{\rm X_0}$ and $0.8\,\lambda_{\rm I}$, respectively. In all configurations the NEUCAL active volume was symmetrically positioned and orthogonal with respect to the beam axis.

The NEUCAL prototype and the testbeam setup have been accurately
modelled with Geant4~\cite{geant4,geant} to validate the results
against the Monte Carlo simulation. A sketch of the geometry of the
configuration c), as implemented in Geant4, is visible in Fig.~\ref{fig:TBGeant}. The Geant4 simulation of NEUCAL has been also cross-checked with Fluka~\cite{fluka1, fluka2} with respect to the single neutron response, as described in~\cite{como, vienna}. 

Testbeam data comparison with Geant4 simulations are done only at the level of the energy deposition in scintillators and on the number of counts in \hetre tubes. In fact, no readout chain response is presently modelled in the simulation.

Two sets of simulated samples have been produced using different
physics lists suitable for the NEUCAL use case~\cite{geantphysref}:
\verb|QGSP_BERT_HP| ({\em QBERT}\/), with Bertini model and
Quark-Gluon String Precompound model to generate the final state for
hadron inelastic scattering respectively below and above $\sim10\GeV$;
\verb|QGSP_BIN_HP| ({\em QBIC}\/), similar to QBERT but
with the Binary Cascade model in place of the Bertini model for the
final state generation below $\sim10\GeV$. Both physics lists
include a high-precision model for low energy neutron
transportation.
The typical size of simulated samples is 20 thousand events for pions
and 80 thousand events for positrons.

\section{Results}

By default data were taken in zero suppression mode, implemented in the
digitizer boards, in order to reduce the event size and the readout time.
The signals up to one millisecond after the initial charged particles
shower were then reconstructed offline to search for delayed neutron
interactions. The neutron signature is an isolated energy deposit in
one of the scintillator modules or a pulse in the \hetre tubes, while
a traversing charged particle will release energy on more than one
scintillator module~\cite{vienna}.

Unfortunately, the photomultipliers and the readout electronics often saturate because of the huge signal induced by the shower core, with different effects seen on data taken with and without zero suppression.
In zero suppression mode the readout hardware spoilt the data in the
time window close to the shower core as a consequence of reflections
seen at the digitizer output lasting for a few hundred nanoseconds.
In the few ten thousand events taken without zero suppression
the board with limited ADC range (V1731) still showed
slowly-recovering saturation effects.
These issues have been understood and fixed in view of the next tests. Eventually, despite the difficult conditions, the samples of Table~\ref{table:samples} have been used to produce the present results.

In the following, all plots report points corresponding to: pion sample data (\includegraphics[height=0.4\baselineskip]{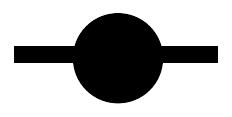}) and pion simulation predictions for QBERT (\includegraphics[height=0.7\baselineskip]{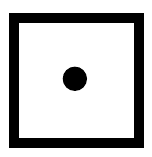}) and QBIC (\includegraphics[height=0.7\baselineskip]{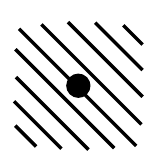}); positron sample data (\includegraphics[height=0.4\baselineskip]{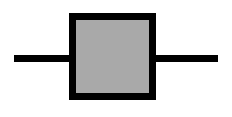}) and positron simulation predictions for QBERT (\includegraphics[height=0.7\baselineskip]{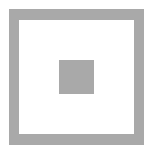}) and QBIC (\includegraphics[height=0.7\baselineskip]{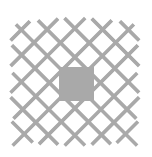}). All plots shown in this paper are normalized to the number
of events which develop a shower in the upstream material: almost all of the positron
and $45\%$ to $60\%$ of pion events, depending on the configuration.

\begin{table}
\caption{\label{table:samples}Testbeam data samples used for each analysis. ZS stands for `zero suppression'.}
\begin{center}
\begin{tabular}{lllll}
\br
Analysis &Time Domain & Mode & $e^+$ sample & $\pi^-$ sample\\
\mr
\hetre counters & $0.1\micros$ -- $0.1\millis$ & ZS        & 39k, $100\GeV$; conf. a) & 76k, $350\GeV$; conf. b)  \\
Scint. `early' & $<1\micros$ & no ZS  & 7k, $150\GeV$; conf. a) & 18k, $350\GeV$; conf. c)  \\
Scint. `late'  & $1\micros$ -- $1\millis$ & ZS  & 39k, $100\GeV$; conf. a) & 76k, $350\GeV$; conf. b)  \\
\br
\end{tabular}
\end{center}
\end{table}

\begin{figure}[b]
\begin{minipage}{0.49\textwidth}
\includegraphics[width=\textwidth]{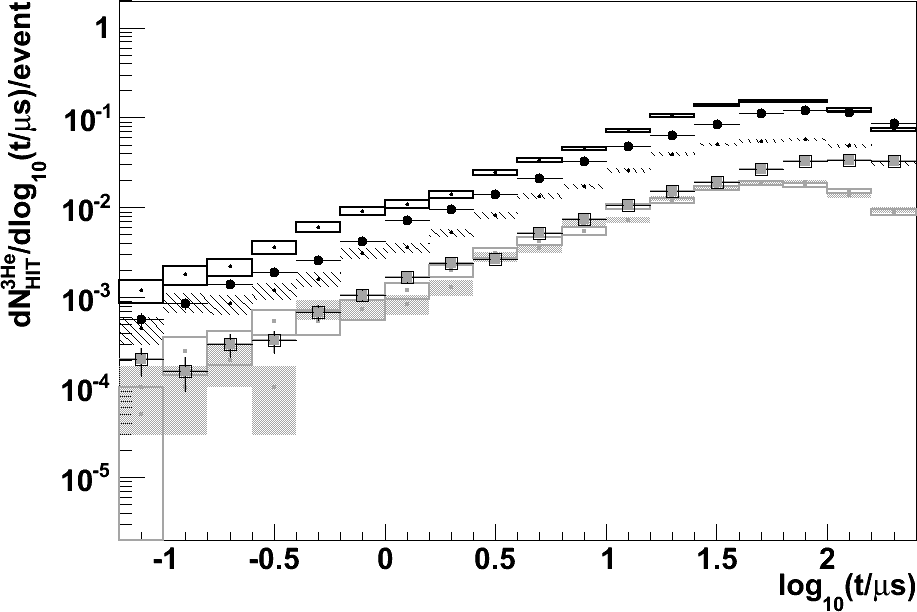}
\caption{\label{fig:3He}Number of pulses per event of all \hetre counters as a function of the logarithm of time, in microseconds.\\}
\end{minipage}
\begin{picture}(0,0) 
\put(-64,-17){\includegraphics[width=0.13\textwidth]{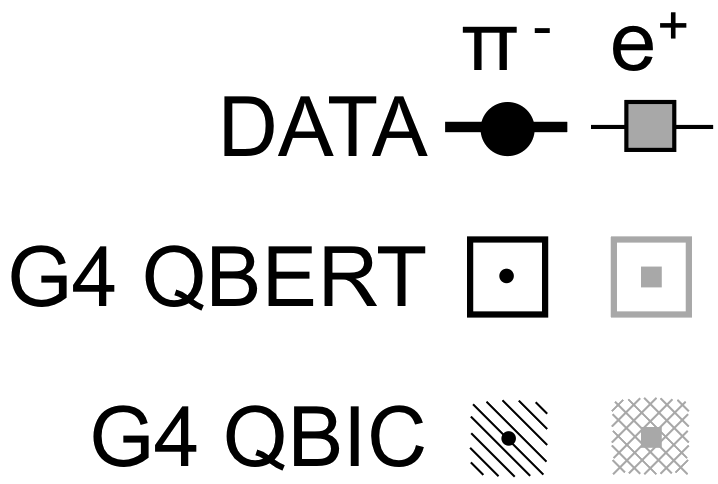}} 
\end{picture}
\hspace{0.01\textwidth}%
\begin{minipage}{0.49\textwidth}
\includegraphics[width=\textwidth]{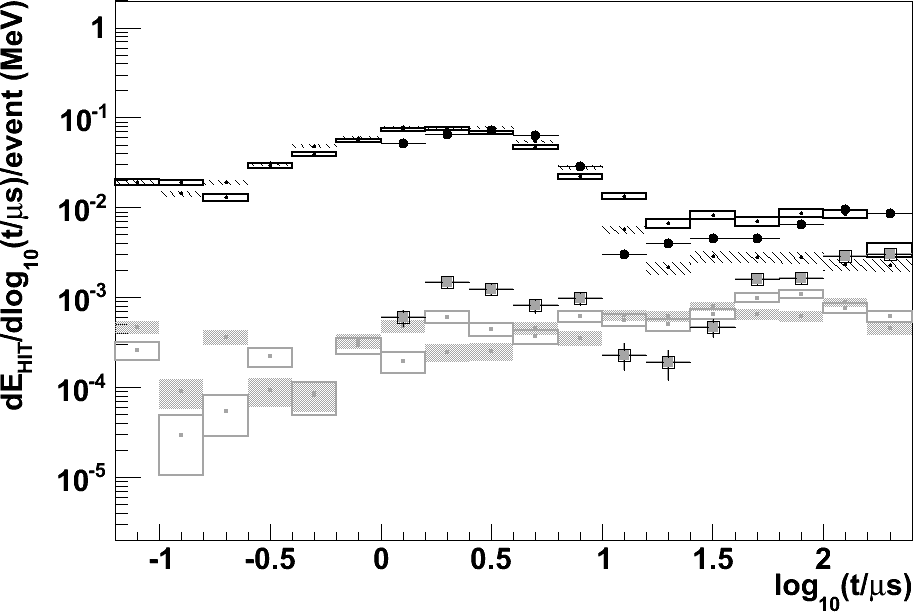}
\caption{\label{fig:late}Hit energy per event of one top-side
  scintillator module
  as a function of the logarithm of time, in microseconds, `late'
  domain.}
\end{minipage} 
\begin{picture}(0,0) 
\put(394,87){\includegraphics[width=0.13\textwidth]{fig/legenda.png}} 
\end{picture}
\end{figure}

The analysis of the traditional \hetre counters (see
Table~\ref{table:samples} for sample details) is an important
validation of the simulation environment. The results are summarized
in Fig.~\ref{fig:3He} where the number of pulses per event on all counters is plotted versus the logarithm of time, in microseconds. Both for data and simulation, the signal time is always referred to the arrival of the shower, taken as time zero.
The behaviour is well reproduced by the simulation, but significant differences in the absolute prediction exist for pions between QBIC and QBERT, observable also in the scintillators analyses reported below. These discrepancies will be investigated further when new datasets from future tests will become available.

The analyses of the scintillator modules data search for neutron interaction candidates ({\em hits}\/) defined as single isolated signals exceeding $0.3\GeV$ in energy. This definition is also implemented at the simulation level. All scintillator modules were previously calibrated in energy with cosmic muons~\cite{tiberio}.
Details of the used samples can be found in
Table~\ref{table:samples}. In particular, the analysis of the `early'
scintillator data ($<1\micros$) is performed on samples without zero
suppression; `late' domain analysis, from $1\micros$ to $1\millis$, relies upon data with zero suppression. 

Data from scintillators are contaminated by several background sources, not modelled in the simulation, for which specific rejection criteria have been deployed. Signal reflections in zero suppression mode are rejected by vetoing specific time delays between subsequent hits; the signals due to off-trigger beam particles can be reduced by asking no hits from different modules in coincidence; spurious effects due to saturation phenomena are reduced by rejecting signals with abnormally long duration.

An example of the results of the `late' analysis, from $1\micros$ to $1\millis$, is given in Fig.~\ref{fig:late} where the hit energy per event is plotted as a function of the logarithm of time, in microseconds, for one top-side module; all modules show similar behaviour.
This range is populated by depositions resulting from neutron captures
on nuclei and it is not directly interesting for the detection of the
neutron moderation. Nevertheless, the good overall agreement between
data and simulation is an indirect indication that neutron flux
estimations are under control. Capture signals could be used as a
complementary handle for neutron detection if they can be made
numerically significant on the single event.

Off-trigger contribution due to beam contamination can be observed on the positron
data for time greater than about $30\micros$ in Fig.~\ref{fig:late},
and also in Fig.~\ref{fig:3He}. This is due to the much greater intensity of the positron beam used in the test.
\begin{figure}[t]
\begin{center}
\includegraphics[width=\textwidth]{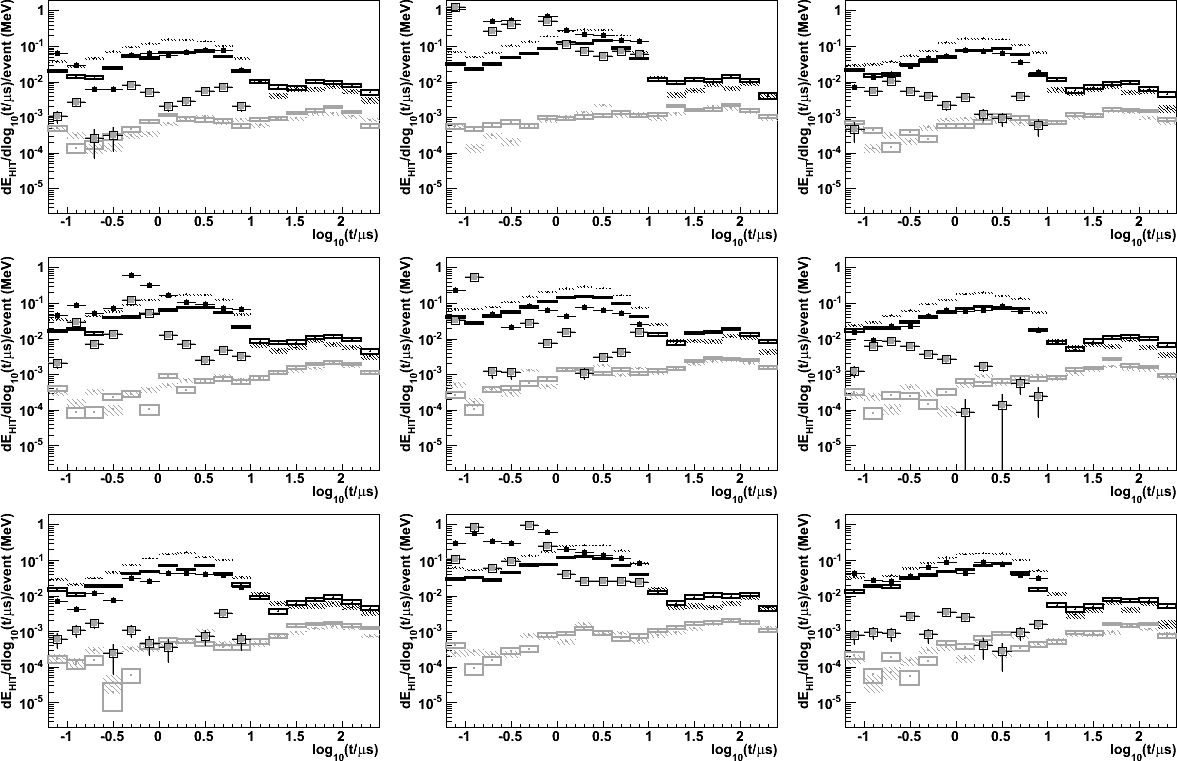}
\end{center}
\caption{\label{fig:early}Hit energy per event as a function of the
  logarithm of time, in microseconds, for positrons and pions from the nine NEUCAL scintillator modules, `early' time domain. Top row corresponds to the upstream module shelf with respect to the beam arrival direction. Same symbol conventions of Fig.~\ref{fig:3He} and Fig.~\ref{fig:late} apply.}
\end{figure}

The `early' time domain analysis ($<1\micros$), performed on data without zero suppression, is represented in Fig.~\ref{fig:early} where the hit energy per event is reported versus the logarithm of time, in microseconds, for all nine modules.
It is evident that central modules suffer the shower core contamination from pions and positrons, especially at short times.
Side modules are less affected by this issue. Nevertheless the agreement between data and simulation is again not satisfactory for positrons, with the exclusion of the two bottom-side modules. This is a saturation effect in the modules read out by the V1731 boards due to the huge signal generated by electromagnetic showers not fully contained by the upstream calorimeter. 

\begin{figure}[t]
\includegraphics[width=0.63\textwidth]{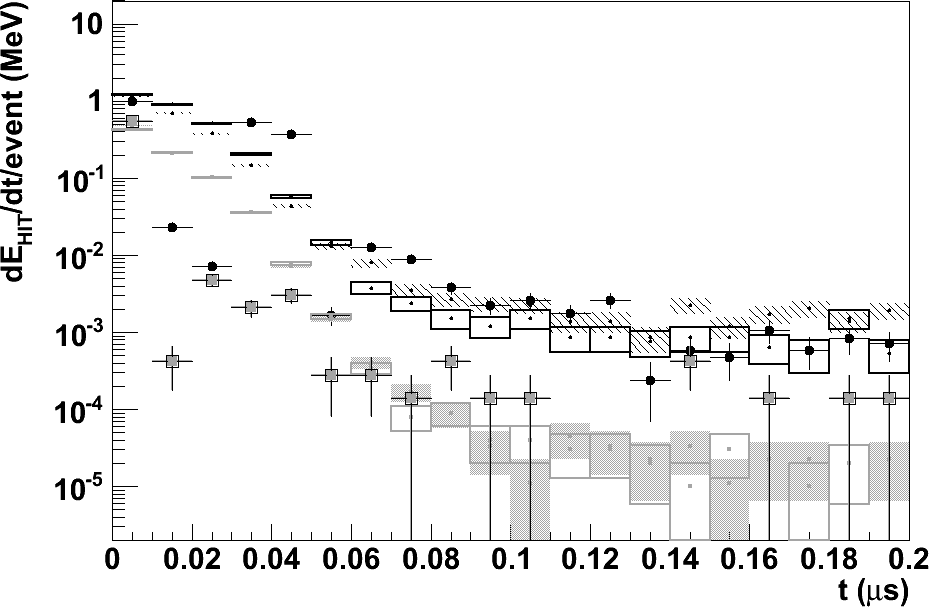}
\begin{picture}(0,0) 
\put(-100,120){\includegraphics[width=0.18\textwidth]{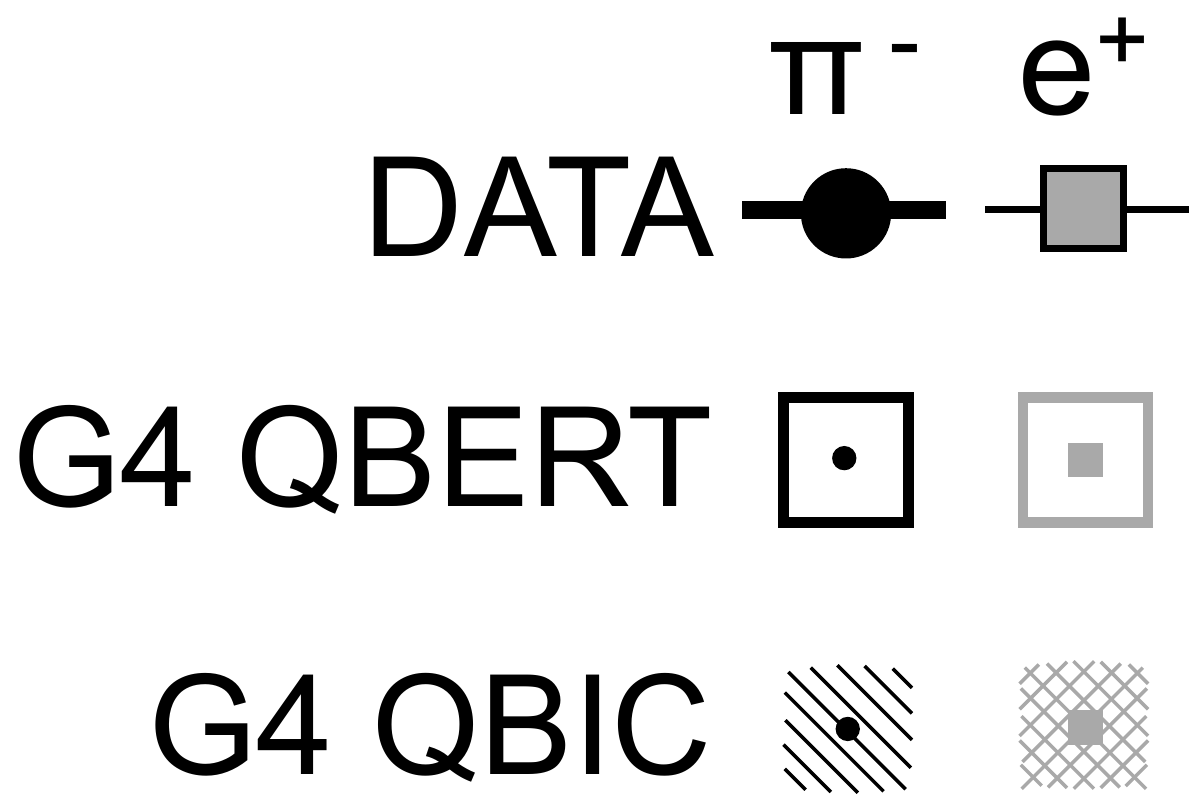}} 
\end{picture}
\hspace{0.01\textwidth}%
\begin{minipage}[b]{0.35\textwidth}\caption{\label{fig:earlyBottomSide}Hit energy per event as a function of time (linear scale) after the shower core.}
\end{minipage}
\end{figure}

The analysis at times before $200\nanos$, where most
of the neutron moderation signal is expected, is thus
restricted to the bottom-side modules that are not directly flooded by
the electromagnetic shower and that are read out by the board V1720,
which has the dynamic range to withstand the huge signal.
The hit energy per event for the bottom-side
modules as a function of time, now represented in linear scale, is shown in Fig.~\ref{fig:earlyBottomSide}.
The agreement with the
simulation is satisfactory with the exception of the region below few
tenth of nanoseconds where a significant shower contamination is still
present and cannot be further reduced due to the limitations of the
present setup. This plot demonstrates that the signal due to neutrons in
pion data is significantly larger than in positron data in the time
interval relevant for the NEUCAL application.

\section{Conclusions}

NEUCAL is a detector designed to exploit the innovative ``active moderation'' technique in the neutron counting. The final aim is the development of a light and compact device, suitable for space experiments, to help electromagnetic calorimeters in hadron to electron discrimination. The NEUCAL prototype has been tested with pion and positron showers, and data compared with an accurate Geant4 simulation. Results demonstrate in general good agreement between data and simulation and are very promising in view of the device application. 
Further confirmations are expected with upcoming tests that will
profit from the experience gained so far and that will be organised with an improved and more performant setup.

\ack
The authors wish to express their gratitude to P~S~Marrocchesi, G~Bigongiari, and P~Maestro
for their support during data taking at CERN and in the analysis of the data.

\section*{References}
\bibliography{neucal}

\end{document}